\documentclass[conference,a4paper]{IEEEtran}

\usepackage[utf8]{inputenc} 
\usepackage[T1]{fontenc}
\usepackage{url}
\usepackage{ifthen}
\usepackage{cite}

\usepackage[cmex10]{amsmath} % Use the [cmex10] option to ensure complicance
\usepackage{graphicx}
\usepackage{algorithm}
\usepackage{algorithmic}

\newtheorem{Example}{Example}
\newtheorem{Theorem}{Theorem}
\newtheorem{Remark}{Remark}
\newtheorem{Lemma}{Lemma}

\newtheorem{Corollary}{Corollary}
\usepackage{subfigure}
\usepackage[english]{babel}
\usepackage{longtable}
\usepackage{tabularx}
\usepackage{amsfonts}
\usepackage{amssymb}
\usepackage{mathrsfs}
\usepackage{ragged2e}
\usepackage{verbatim}

\begin{document}
\title{On DNA Codes using the Ring $\mathbb{Z}_{4} + w\mathbb{Z}_{4}$}
\author{
 \IEEEauthorblockN{Dixita Limbachiya  \IEEEauthorrefmark{1},
                      Krishna Gopal Benerjee \IEEEauthorrefmark{1},
                     Bansari Rao\IEEEauthorrefmark{2}
                     and Manish K. Gupta\IEEEauthorrefmark{1}}
   \IEEEauthorblockA{\IEEEauthorrefmark{1}%
                     Dhirubhai Ambani Institute of Information and Communication Technology\\
					  Email: dlimbachiya@acm.org, kg.benerjee@gmail.com and mankg@computer.org  \IEEEauthorrefmark{1}             
                     %\{dlimbachiya@acm.org, kg.benerjee@gmail.com, mankg@computer.org\}
                     }
   \IEEEauthorblockA{\IEEEauthorrefmark{2}%
                    University of Florida, Florida, USA.
					Email: bansarirao29@ufl.edu                    
                    }
 }

\maketitle

%%%%%%
%% Abstract: 
%% If your paper is eligible for the student paper award, please add
%% the comment "THIS PAPER IS ELIGIBLE FOR THE STUDENT PAPER
%% AWARD." as a first line in the abstract. 
%% For the final version of the accepted paper, please do not forget
%% to remove this comment!
%%
\begin{abstract}
In this work, we study the DNA codes from the ring $R$ = $\mathbb{Z}_4 + w\mathbb{Z}_4$, where $w^2 = 2+2w$ with $16$ elements. We establish a one to one correspondence between the elements of the ring $R$ and all the DNA codewords of length 2 by defining a distance preserving Gau map $\phi$. Using this map, we give several new classes of the DNA codes which satisfies reverse and reverse complement constraints. Some of the constructed DNA codes are optimal.
\end{abstract}

%% The paper must be self-contained. However, if you are referring to
%% a full version for checking certain proofs, please provide the
%% publically accessible location below.  If the paper is completely
%% self-contained, you can remove the following line from your
%% submission.

%\textit{A full version of this paper is accessible at: \url{http://www.isit2018.org/}}

\section{Introduction}\label{intro}
DNA (DeoxyriboNucleic Acid) is the basic programming unit of life. DNA consists of four building blocks called nucleotides viz. Adenine ($A$), Thymine ($T$), Cytosine ($C$) and Guanine ($G$). DNA forms a double helix by the process called hybridization in which each nucleotide on a DNA strand is held by the hydrogen bonds with its complementary base, where $A$ is complement to $T$, $G$ is complement to $C$. The use of the DNA for computation was first demonstrated in an experiment by L. Adleman \cite{adleman1994molecular} in 1994 by solving an instance of Hamilton path problem using the DNA strands. His method used the idea of DNA hybridization which is the backbone for any computation using the DNA strands. But it is also the source of errors. Errors in the DNA computing can be avoided by constructing the DNA strands with a particular set of constraints \cite{milenkovic2005dna} that makes the DNA strands feasible for the practical implementation. These set of DNA strands (DNA codes) which are sufficiently dissimilar have been designed by using various computational and theoretical approaches in the literature \cite{limbachiya2016art}. Moreover, coding theoretic approaches \cite{gaborit2005linear} have been used to construct the DNA codes from codes over fields and rings. This paper uses one such approach to develop the DNA codes from the ring $R = \mathbb{Z}_{4} + w\mathbb{Z}_{4}$, where $w^2=2+2w$.

%$\textit{ie}$. For $\textbf{x}, \textbf{y} \in \mathscr{C}$, $d_{H}(\textbf{x}$, $\textbf{y}) = \# \{i $ $ |$ $ x_i \neq y_i\}$.

Let $\Sigma_{DNA} =\{A,C,G,T\}$. The DNA code is the set of $M$ distinct DNA strands each of length $n$ such that the Hamming distance between any two DNA strands is at least $d_{H}$, where, $d_{H}$ = $\min\{d_H(\textbf{x},\textbf{y}):\textbf{x}\neq\textbf{y},\ \forall\ \textbf{x},\textbf{y}\in\mathscr{C}_{DNA}\}$. Mathematically, the DNA code is denoted by $\mathscr{C}_{DNA}(n,M,d_{H}) \subseteq \Sigma^{n}_{DNA} =\{A,T,G,C\}^n$. The DNA strands are also called as DNA codewords.

%with each codeword of length $n$, number of DNA codewords $M$ and minimum Hamming distance \textbf{$d_{H}$}. The Hamming distance $d_{H}(\textbf{x}$, $\textbf{y})$ between two codewords $\textbf{x}$ and $\textbf{y}$ is the number of positions in which $\textbf{x}$ and $\textbf{y}$ differs. Note that a Minimum Hamming distance 

\begin{Example} $\mathscr{C}_{DNA}$ =  $\{AGAG$, $AGGA$, $AGCT$, $AGTC$, $GAAG$, $GAGA$, $GACT$, $GATC$, $CTAG$, $CTGA$, $CTCT$, $CTTC$, $TCAG$, $TCGA$, $TCCT$, $TCTC\}$ is a DNA code  $\mathscr{C}_{DNA}$ with parameters $n=4$, $M = 16$ and $d_{H}=2$.
\label{eg}
\end{Example} 

Generally, the dissimilarity measure used for the DNA codewords is defined by the Hamming distance between the DNA codewords. But to avoid the mis-hybridization between the set of DNA codewords, it is important to consider the distance between the given DNA codeword, its reverse and reverse complement DNA codewords. This motivates researchers to define the combinatorial constraints on the DNA codewords.

%These DNA codewords with desired properties that enables the DNA codewords to perform the computation are characterized by DNA constraints. 
For the given DNA codeword $\textbf{x} = (x_1 \ x_2 \  \ldots \ x_{n-1} \ x_n)$, the reverse of the DNA codeword is defined as $\textbf{x}^{\textbf{r}} =(x_n \ x_{n-1} \ \ldots \ x_2 \ x_1)$, the Watson - Crick complement or simply complement of the DNA codeword is defined as $\textbf{x}^{\textbf{c}}  = (x_1^c \ x_2^c \ \ldots \ x_{n-1}^c \ x_n^c)$, the reverse complement of the DNA codeword is defined as $\textbf{x}^{\textbf{rc}} = (x_n^c \ x_{n-1}^c \ \ldots \ x_2^c \ x_1^c)$, where each $x_i \in \{A,T,G,C\}$ for $i=1,2,\ldots n$ and $A^c = T$, $T^c = A$, $G^c = C$, $C^c = G$. For Example, if $\textbf{x}$ = $AACT$ then $\textbf{x}^\textbf{r}$ = $TCAA$, $\textbf{x}^\textbf{rc}$ = $AGTT$. These constraints ensures that the DNA hybridize with the correct match and avoids errors in the computation \cite{marathe2001combinatorial,king2003bounds}. A DNA code with some predefined distance $d$ should satisfy one or all the combinatorial constraints which are the Hamming distance, Reverse, Reverse complement and GC content constraints. 

The DNA code given in the Example \ref{eg}, satisfies all the combinatorial constraints. For any given $\textbf{x}$, $\textbf{y}$ $\in \mathscr{C}_{DNA}$, note that $d_H(\textbf{x}^\textbf{r},\textbf{y}^\textbf{c})$ =  $d_H(\textbf{x}^\textbf{rc},\textbf{y})$ =  $d_H(\textbf{x},\textbf{y}^\textbf{rc})$ =  $d_H(\textbf{x}^\textbf{c},\textbf{y}^\textbf{r})$. %because $d_H(\textbf{x}^\textbf{r},\textbf{y}^\textbf{c})\geq 2=d_{H}$ for any $\textbf{x},\textbf{y}\in\mathscr{C}_{DNA}$, where $\textbf{x}^\textbf{r}\neq\textbf{y}^\textbf{c}$. 

The progress of DNA codes from an algebraic coding has ignited the interest of coding theorist to develop the DNA codes using rings. In 2001, reversible cyclic DNA codes using the quaternary alphabets was first developed in \cite{rykov2001dna}. But it considered only the reverse constraint for the DNA codes. King et al., in the year 2005, constructed the linear DNA codes from the ring $\mathbb{Z}_4$ \cite{gaborit2005linear} that satisfies reverse and reverse complement constraints. In order to obtain good DNA codes, researchers used different ring structures. Cyclic codes from the finite ring $\mathbb{F}_2 + u\mathbb{F}_2$ were obtained using similarity measures (a special kind of distance similarity) instead of Hamming distance \cite{siap2008similarity}. In \cite{siap2009cyclic}, I. Siap et al. introduced the cyclic DNA codes from the ring $\mathbb{F}_2 + u\mathbb{F}_2$, where $u^2=1$ based on the deletion distance. By using the same ring $\mathbb{F}_2 + u\mathbb{F}_2$ with $u^{2} =0$, the DNA cyclic codes of an even length was constructed by Liang and Wang in \cite{liangcyclic}. DNA cyclic irreversible odd length codes of the type simplex and Reed Muller codes from the ring $\mathbb{F}_2 + u\mathbb{F}_2$ were studied in \cite{guenda2013construction}. Odd length DNA codes was given which satisfies the Hamming distance constraints from the commutative ring $\mathbb{F}_2[u]/<u^4-1>$ with $u^4 = 1$ in \cite{guenda2013cyclic}. A new ring $\mathbb{Z}_{4} + u\mathbb{Z}_{4}$ with $16$ elements was introduced by Yildiz et al. in \cite{yildiz2014linear} and DNA codes of odd lengths from the ring were studied in \cite{pattanayak2015cyclic}.

Recently, a non-commutative ring was used for designing the reversible DNA codes from the skew cyclic codes. Generalized non-chain ring $\mathbb{F}_{16}+ u\mathbb{F}_ {16}+ v\mathbb{F}_ {16}+ uv\mathbb{F}_ {16}$ was used to construct the reversible DNA codes in \cite{gursoy2017reversiblegene}. Very recently, Oztas et al. gave a novel approach for constructing the reversible DNA codes using the ring $\mathbb{F}_2[u]/(u^{2k}- 1)$ \cite{oztas2017novel}. For more background on this, the reader is referred to the further references in the literature \cite{limbachiya2016art,zhucyclic}.  

% DNA codes over general ring $\mathbb{F}_{16}$ using lifted polynomial was studied in \cite{oztas2013lifted}. Bayram et al. \cite{bayram2015codes} constructed DNA codes using skew constacyclic over the ring $\mathbb{F}_4+v\mathbb{F}_4$. A direct map between $64$ elements of the ring to $64$ DNA codons (three nucleotides) DNA codes was given in  \cite{bennenni2015new} over the ring $R = \mathbb{F}_2[u]/(u^6)$.  oztas2017dna The ring $\mathbb{F}_4 [u] /<u^2+1>$ with $u^2 =1$ was used for DNA codes of the length $6$ in \cite{ma2015cyclic}. yildiz2012cyclic  \cite{gursoy2017reversible}

Choie and Dougherty studied the properties of the ring $R = \mathbb{Z}_4 + w\mathbb{Z}_4$ for the first time in \cite{choie2005codes}. Using the structure of the ring, the DNA cyclic codes of odd lengths from the ring $\mathbb{Z}_4 + w\mathbb{Z}_4$, where $w^2 = 2$ was proposed in \cite{dertli2016cyclic}. 

Motivated by this, we investigate the structure of the ring $R = \mathbb{Z}_4 + w\mathbb{Z}_4$ from \cite{choie2005codes} and use it to construct the DNA codes of even lengths. In this paper, the ring $\mathbb{Z}_{4} + w\mathbb{Z}_{4}$, where $w^2=2+2w$ with $16$ elements is considered. A correspondence between the ring elements and the DNA codewords of length $2$ is defined via a distance preserving Gau map $\phi$. We propose a new type of distance called the Gau distance on the ring $R$. Properties of the distance with respect to the DNA codes are explored. Several new families of the DNA codes are obtained which satisfies the Hamming, reverse and reverse complement constraints. In order to obtain the DNA codes with a fixed GC content, the DNA codewords that violate the $GC$ content constraint are removed.

%Contributions of this work is, the authors have constructed DNA codes over ring $R$ = $\mathbb{Z}_4 + w\mathbb{Z}_4$ with $w^2 = 2+2w$. Gau map is defined from elements of ring to DNA nucleotides of length 2. Family of codes like simplex type codes, Reed Muller codes and Octacodes are used to design the DNA codes of different even length $n$, size $M$ and minimum distance $d$.% For each family of code, the generator matrix over the ring $R$ is defined and DNA codes are developed by using defined mapping. 

The paper is organized as follows. Section 2 gives a brief background of the ring $R$ = $\mathbb{Z}_{4} + w\mathbb{Z}_{4}$. Section 3 discusses the Gau distance, the Gau map and the DNA codes constructed from the ring $R$. Section 4 illustrates the results on the families of the reverse and reverse complement DNA codes with Examples. Section 5 the $r^{th}$ order Reed Muller Type code. Section 6 concludes the paper with some general Remarks.

%%%%%%
%% Appendix:
%% If needed a single appendix is created by
%%
%\appendix
%%
%% If several appendices are needed, then the command
%%
% \appendices
%%
%% in combination with further \section-commands can be used.
%%%%%%

\section{Preliminaries on the Ring $\mathbb{Z}_{4} + w\mathbb{Z}_{4}$}\label{sec:2}
The structure of finite commutative local chain ring $R$ = $\mathbb{Z}_{4} + w\mathbb{Z}_{4}$ = $\{a+bw: a,b\in\mathbb{Z}_4\mbox{ and }w^2=2+2w\}$ is discussed here. %\cite{bandi2015self}. 
The ring $R$ is a principal ideal ring with $16$ elements. For the ring $R$, the set of zero divisors is $\{a+bw: a\in 2\mathbb{Z}_4,b\in \mathbb{Z}_4\}$ = $\{0,2,w,2+w,2w,2+2w,3w,2+3w\}$ and the set of units is $\{a+bw: a\in 2\mathbb{Z}_4+1,b\in \mathbb{Z}_4\}$ = $\{1,3,1+w,3+w,1+2w,3+2w,1+3w,3+3w\}$. The ring $R$ has $5$ distinct ideals as $\big \langle 0\big \rangle\subset\big \langle 2w\big \rangle\subset\big \langle 2\big \rangle = \big \langle 2+2w\big \rangle\subset\big \langle w\big \rangle = \big \langle 2+w\big \rangle = \big \langle 3w\big \rangle = \big \langle 2+3w\big \rangle \subset R$.
%\begin{center}
%$\begin{array}{ll}
%\big \langle 0\big \rangle = \{0\}  & \\
%\big \langle 2w\big \rangle = \{0,2w\} \\ 
%\big \langle 2\big \rangle = \{0,2,2w,2+2w\}= \big \langle 2+2w\big \rangle & \\ 
%\big \langle w\big \rangle  = \{0,2,w,2+w,2w,2+2w,3w,2+3w\}
%                             = \big \langle 2+w\big \rangle = \big \langle 3w\big \rangle = \big \langle 2+3w\big \rangle \\ 
%% \big \langle 1\big \rangle  = R = \big \langle 3\big \rangle = \big \langle 1+w\big \rangle = \big \langle 3+w\big \rangle = \big \langle 1+2w\big \rangle = \big \langle 3+2w\big \rangle = \big \langle 1+3w\big \rangle = \big \langle 3+3w\big \rangle  \\
%\big \langle x\big \rangle  = R\ (x = 1, 3, 1+w, 3+w, 1+2w, 3+2w, 1+3w, 3+3w) \\
%\end{array}$
%\end{center}
%Note that $\big \langle 0\big \rangle\subset\big \langle 2w\big \rangle\subset\big \langle 2\big \rangle = \big \langle 2+2w\big \rangle\subset\big \langle w\big \rangle = \big \langle 2+w\big \rangle = \big \langle 3w\big \rangle = \big \langle 2+3w\big \rangle\subset R$.

Any subset of $R^n$ is called the code over the ring $R$. Any sub-module of $R^n$ is called the linear code $\mathscr{C}$ over the ring $R$. The standard form of the generator matrix $G$ of the linear code $\mathscr{C}$ over the ring $R=\mathbb{Z}_4 + w\mathbb{Z}_4$ is given by:
\begin{equation}\label{general_formR}
G=\left(
\begin{array}{ccccc}
I_{k_0} & A_{0,1}   & A_{0,2}   & A_{0,3}   & A_{0,4}  \\
0       & wI_{k_1} & wA_{1,2} & wA_{1,3} & wA_{1,4} \\
0       & 0         & 2I_{k_2}  & 2A_{2,3}  & 2A_{2,4} \\
0       & 0         & 0         & 2wI_{k_3}  & 2wA_{3,4} \\
\end{array} \right),
\end{equation} 
where the matrices $A_{i,j}$ are defined over the ring $R$. Note that $A_{i,j} = B_{i,j}^{1} + wB_{i,j}^{2} + 2B_{i,j}^{3} + 2wB_{i,j}^{4}$, where $B_{i,j}^{k}$ is binary matrix for $0\leq i<j\leq 4$ and $k=1,2,3,4$. A code with a generator matrix
in this form is of type $\{k_0, k_1, k_2, k_3\}$ and has $16^{k_0}8^{k_1}4^{k_2}2^{k_3}$ codewords \cite{choie2005codes}. We denote the row span of the matrix $G$ on the ring $R$ by $<G>_R$.

In the next Section, we give a mechanism via the Gau map to construct the DNA codes from the ring $R$  that satisfies the Hamming distance, reverse and reverse complement constraints. 

%%%%%%%%%%%%%%%%%%%%%%%%%%%%%%%%%%%%%%%%%%%%%%%%%%%%%%%%%%%%%%%%%%%%%%%%%%%%%%%%%%%%%%%%%%%%%%%%%
\section{DNA Codes using the Ring $R$ = $\mathbb{Z}_4+w\mathbb{Z}_4$}\label{codesoverR}
In order to construct the DNA codes using the ring $R$, a correspondence between the elements of the ring $R$ and the DNA alphabets is required. 
We give an isometry (distance preserving map) between the codes over the ring and the DNA codes. 
To define a distance on the ring $R$ (hence eventually on $R^n$), the elements of the ring and the DNA alphabets can be arranged in a manner (see the Matrix $\mathscr{M}$ in the equation \ref{gaumatrix}) such that the Hamming distance $d_H$ between any two distinct pair of DNA nucleotides in the same row or same column is 1, otherwise it is 2. This motivates us to define a distance called the Gau distance on the elements of the ring $R$ such that this property is preserved. 

\begin{equation}\label{gaumatrix}
\begin{array}{cc}
    \mathscr{M} = &  \begin{array}{cc}
         & 
\begin{array}{cccccccccccccccccccc}
            A & & & G & & & C & & & T
        \end{array}  
        \\
        \begin{array}{c}
           A \\
            G \\
            C \\
            T 
        \end{array}  
        &
        \left(\begin{array}{cccc}
         0 & 1 & 2+3w & 3+3w  \\
        3 & 2 & 1+3w & 3w     \\
        2+w & 3+w & 2w & 1+2w \\
        1+w & w & 3+2w & 2+2w
\end{array}\right)
    \end{array}
\end{array}
\end{equation}

For $x, y \in R$, let $x = m_{i,j} \in \mathscr{M}$, $y = m_{i^{'},j^{'}} \in \mathscr{M}$ for some $0 \leq i,j \leq 3$ and $0 \leq i^{'},j^{'} \leq 3 $ then, the Gau distance $d_{Gau}$ can be defined as \\ \begin{equation} d_{Gau}(x, y) = min\{1, i + 3i^{'}\} + min\{1, j + 3j^{'}\}.
\end{equation}
%let $\mathscr{M}(x) = (x_i,x_j)$ and $\mathscr{M}(y) = (y_i,y_j)$ be the indexes of the elements $x$ and $y$ of the matrix $\mathscr{M}$ 

\begin{Example}
For $x=2$, $i=1, \ j=1$ and $y= 2+2w$, $i^{'}=3, \ j^{'}=3$, the Gau distance $d_{Gau}(2, 2+2w) = 2$.
\label{distmapeg}
\end{Example}
 It is easy to see that the Gau distance $d_{Gau}$ is indeed a metric on the elements of the ring $R$. Using the set of zero divisors 
 $Z$ = $\{0,2,w,2+w,2w,2+2w,3w,2+3w\}$ and set of units  $U$ = $\{1,3,1+w,3+w,1+2w,3+2w,1+3w,3+3w\}$ of R, one can simplify the formula for the Gau distance $d_{Gau}$. 
 For both $x$ and $y$ $\in Z$ or both $x$ and $y$ $\in U$ we have, 
%\[
\begin{center}
$d_{Gau}(x,y) = \left\{ \begin{array}{ll}
0 & $ $ if $ $ x=y , \\
1 & $ $ if $ $x \neq y$ and $x+3y \in \{2+w,2+3w\}, \\
2 &  $ $ otherwise. %if\ x+3y = 1+2w\mbox{ or }3+2w\pmod4.  % 2,2w,2+2w,2+w,2+3w,
\end{array}\right.$
\end{center}
%\]

For any $x\in\{0,2,2w,2+2w\}$ and $y\in U$, 
\begin{center}
$d_{Gau}(x,y) = \left\{ \begin{array}{ll}
0 & if $ $  x=y , \\
1 & if $ $  x \neq y$ and $x+3y \in \{1,3,1+w,3+3w\}, \\
2 & if $ $  x \neq y$ and $x+3y \in \{1+3w,3+w,1+2w,3+2w\}.
\end{array}\right.$
\end{center}

For any $x\in\{w,3w,2+w,2+3w\}$ and $y\in U$,
\begin{center}
$d_{Gau}(x,y) = \left\{ \begin{array}{ll}
0 & if $ $  x=y , \\
1 & if $ $  x \neq y$ and $x+3y \in \{1,3,3+w,1+3w\}, \\
2 & if $ $  x \neq y$ and $x+3y \in \{3+3w,1+w,1+2w,3+2w\}.
\end{array}\right.$
\end{center}

For any two arbitrary vectors $\textbf{x} =(x_1 \  x_2 \  x_3 \ \ldots \ x_n) \in R^n$ and $\textbf{y} =(y_1\ y_2\ y_3\ \ldots \ y_n)$ $ \in R^n$, the Gau distance $d_{Gau}(\textbf{x}, \textbf{y}) = \sum_{i=1}^n d_{Gau}(x_i, y_i)$ is a metric on $R^n$ induced by the metric on the elements of the ring R. We use the same notation $d_{Gau}$ for both the metrics on $R$ and $R^n$. For a linear code $\mathscr{C}$ on $R$, one can define a minimum Gau distance  $d_{Gau}$ = $min\{d_{Gau}(\textbf{x}, \textbf{y}): \textbf{x}, \textbf{y} \in \mathscr{C}$ and $x \neq y \}$. 

\begin{Example}
For $\textbf{x}= (2 \ 2+2w\ 0 \ 2w)$ and $\textbf{y}= (0\ 2 \ 2w \ 2+2w)$, $d_{Gau}(\textbf{x}, \textbf{y}) = 8$.
\end{Example}

Now, we define a Gau map $\phi$ (as shown in the Table \ref{gaumapping}) from the elements of $R$ to all DNA vectors of length $2$ as:

\begin{equation}\label{dnamap}
\phi: R \rightarrow  \Sigma_{DNA}^2
\end{equation}
%this is the final mapping
\begin{center}
\begin{table}
\centering
\begin{tabular}{|c|c||c|c|}
\hline
Ring element & DNA image &   Ring element & DNA image \\ 
$x$    &  $\phi(x)$  & $x$     & $\phi(x)$   \\
\hline \hline
$0$   & $AA$        & $1$  & $AG$  \\ 
\hline
$2$  & $GG$ &  $3$ & $GA$ \\
\hline
$w$     &  $TG$       & $1+w$  & $TA$       \\
\hline
$2+w$   & $CA$ & $3+w$ & $CG$ \\ 
\hline
$2w$     &  $CC$       & $1+2w$     & $CT$  \\ 
\hline
$2+2w$  & $TT$ & $3+2w$ & $TC$ \\
\hline
 $3w$    &  $GT$       & $1+3w$    & $GC$  \\ 
\hline
$2+3w$    & $AC$ & $3+3w$ & $AT$ \\
\hline
\end{tabular}
\caption{A bijective mapping $\phi$: $R\rightarrow \Sigma_{DNA}^2$ is illustrated. For each $x \in R$, $\phi(x)^c$ and $\phi(x)^r$ denotes the complement and reverse codeword of the DNA codeword $\phi(x)$. Note that this map $\phi$ preserves the complement and reverse constraints by the relations $\phi^{-1}(\phi(x)^c)$ = $x + (2+2w)$ and $x$ + $\phi^{-1}(\phi(x)^r)$ = $0$ $\forall$ $x \in R$.}
\label{gaumapping}
\end{table}
\end{center}
\vspace{-0.7cm}
One can observe the following properties of the Gau map $\phi$.

\begin{enumerate}
%item There exists a one-to-one correspondence between the DNA codewords of length $2$ and the elements of the ring $R$.
\item The additive inverse of each element $x \in R$ is unique, similarly the reverse $\phi(x)^r$ of each $\phi(x) \in \Sigma_{DNA}^2$ is unique.
%\item For $x \in R, $ $ \exists$ a unique additive inverse, similarly for each $\phi(x) \in \Sigma_{DNA}^2, $ $\exists$ a unique reverse DNA $\phi(x)^r$.
\item Four elements $0,2,2w,2+2w \in R$ are self invertible under the addition operation. One can observe that the DNA nucleotides, $\phi(0) = AA, \phi(2) = GG, \phi(2w) = CC, \phi(2+2w) = TT$ are also self reversible.% from the Table \ref{gaumapping}.
%\item For the ring $R$, $\exists$ four distinct elements $0,2,2w,2+2w \in R$ which are self invertible under the addition operation. From Table \ref{gaumapping}, one can observe that, $\phi(0) = AA, \phi(2) = GG, \phi(2w) = CC, \phi(2+2w) = TT$ are the self reversible are mapped self additive invertible elements of the ring $R$.
\item For each $\phi(x)\in \Sigma_{DNA}^2$,  $\exists$  $\phi(y) \neq \phi(x)$ such that $\phi(x)^c = \phi(y)$ and $\phi(y)^c = \phi(x)$. Similarly, for any $x\in R$ there exists $y\in R$ ($y \neq x$) such that $y = x+a = a+x$ and  $x = y+a = a+y$, for some $a\in R$. In this work, we have consider $a = 2+2w$.
\item The Gau map $\phi$ has a property $\phi^{-1}(\phi(x)^c)$ = $x + (2+2w)$ and $x$ + $\phi^{-1}(\phi(x)^r)$ = $0$ $\forall$ $x \in R$. %which ensures that reverse and reverse complement constraints are satisfied by the DNA codes.
\item For the ring $R$, $\exists$ four distinct elements $3+3w, 1+w, 3+w, 1+3w \in R$ such that $x+a$ is additive inverse of $x$, where $a=2+2w$. From the Table \ref{gaumapping}, one can observe that $\phi(x)^r = \phi(x)^c$ for $x\in\{3+3w, 1+w, 3+w, 1+3w\}$. 
%some $\phi(x) \in \Sigma_{DNA}^2 $. For $ x = 3+3w, \phi(x) = AT, \phi(x)^r = TA = \phi(x)^c$. In the same manner it is true for $\phi(1+w) = TA,  \phi(3+w) = CG, \phi(1+3w) = GC$.
\end{enumerate}

For any $\textbf{x}$ = $(x_1\ x_2\ \ldots \ x_{n-1} \ x_n)\in R^n$, we define 
\newline $\phi(\textbf{x}) = (\phi(x_1)\ \phi(x_2)\ \ldots \phi(x_{n-1})\ \phi(x_n))\in\sum_{DNA}^{2n}$, 
\newline  $\phi^{-1}(\phi(\textbf{x})) = (\phi^{-1}(\phi(x_1))\ldots\phi^{-1}(\phi(x_n)))\in R^n$, 
\newline $\phi^{-1}(\phi(\textbf{x})^r)= (\phi^{-1}(\phi(x_n)^r)\ldots$ $\phi^{-1}(\phi(x_1)^r))\in R^n $ and  
\newline $\phi^{-1}(\phi(\textbf{x})^c) = (\phi^{-1}(\phi(x_1)^c)\ldots\phi^{-1}(\phi(x_{n})^c ))\in R^n$.

%\[
%\begin{array}{rl}
%     \phi(\textbf{x}) & = (\phi(x_1)\ \phi(x_2)\ \ldots \phi(x_{n-1})\ \phi(x_n))\in\sum_{DNA}^{2n}   \\
%     \phi^{-1}(\phi(\textbf{x})) & = (\phi^{-1}(\phi(x_1))\ \phi^{-1}(\phi(x_2))\ \ldots \phi^{-1}(\phi(x_{n-1}))\ \phi^{-1}(\phi(x_n)))\in R^n \\ 
%     \phi^{-1}(\phi(\textbf{x})^r) & = (\phi^{-1}(\phi(x_n)^r)\ \phi^{-1}(\phi(x_{n-1})^r)\ \ldots$ $\phi^{-1}(\phi(x_1)^r))\in R^n \mbox{ and } \\
%     \phi^{-1}(\phi(\textbf{x})^c) & = (\phi^{-1}(\phi(x_1)^c)$ $\phi^{-1}(\phi(x_2)^c)$ $\ldots$ $\phi^{-1}(\phi(x_{n-1}^c) \phi^{-1}(\phi(x_{n})^c ))\in R^n.
%\end{array}
%\]

For any subset $\mathscr{C} \subseteq R^n$, the DNA code $\phi(\mathscr{C}) = \{\phi(\textbf{x}): \ \forall \ \textbf{x} \in \mathscr{C}\}\subset\sum_{DNA}^{2n}$. 

\begin{Theorem}\label{distanceisometry}%isomerty\textbf{need to write proof}
$\phi: (R^n, d_{Gau})$ to $(\Sigma_{DNA}^{2n},d_H)$ is a distance preserving map.
\end{Theorem}

\begin{Example}
For $\textbf{x}= (2 \ 0 \ 2+2w \ 2)$, $\textbf{y}= (2+2w \ 0 \ 2w \ 2)$, $\phi(\textbf{x}) = (GG \ AA \ TT \ GG)$, $\phi(\textbf{y}) = (TT\ AA\ CC\ GG)$, $d_{Gau}(\textbf{x}, \textbf{y})$ = $4$, one can observe that $d_{H}(\phi(\textbf{x}),\phi(\textbf{y})) = 4$.
\label{distmapeg2}
\end{Example}

\begin{Remark}\label{closedrev}
For any $\textbf{x} \in \mathscr{C}$, $\phi^{-1}(\phi(\textbf{x})^r) \in  \mathscr{C}$ if and only if the DNA code $\phi(\mathscr{C})$ is closed under reverse. % $\phi(\textbf{x})^r \in \mathscr{C}_{DNA}$ $\forall$ $\phi(\textbf{x}) \in \mathscr{C}_{DNA}$ then the DNA code $\mathscr{C}_{DNA}$ is closed under reverse.
\end{Remark}

\begin{Remark}\label{closedcomp}
For any $\textbf{x} \in \mathscr{C}$, $\phi^{-1}(\phi(\textbf{x})^c) \in  \mathscr{C}$ if and only if the DNA code $\phi(\mathscr{C})$ is closed under complement constraint. %$\phi(\textbf{x})^c \in \phi(\mathscr{C})$ $\forall$ $\phi(\textbf{x}) \in \mathscr{C}_{DNA}$ then the DNA code $\mathscr{C}_{DNA}$ is closed under complement.
\end{Remark}

\begin{Remark}
For any $\textbf{x} \in \phi(\mathscr{C})$, if $\textbf{x}^r \in \phi(\mathscr{C})$ and $\textbf{x}^c \in \phi(\mathscr{C})$ then $\textbf{x}^{rc} \in \phi(\mathscr{C}) $. %where $\mathscr{C}$ is a code over the ring $R$.
\label{r_rc_exist}
\end{Remark}

% \begin{remark}
% For any $\textbf{x}, \textbf{y}\in R^n$, $d_H(\phi(\textbf{x}),\phi(\textbf{y})^r) = d_H(\phi(\textbf{x})^r,\phi(\textbf{y}))$, $d_H(\phi(\textbf{x}),\phi(\textbf{y})) = d_H(\phi(\textbf{x})^c,\phi(\textbf{y})^c)$ and $d_H(\phi(\textbf{x})^c,\phi(\textbf{x})^r) = d_H(\phi(\textbf{x}),\phi(\textbf{x})^{rc})$. 
% \end{remark} 

\begin{Lemma}%[Reverse]
For any $\textbf{x}, \textbf{y}  \in R^n, \phi^{-1}(\phi(a\textbf{x}+b\textbf{y})^r) = a\phi^{-1}(\phi(\textbf{x})^r) + b\phi^{-1}(\phi(\textbf{y})^r)$, where $a,b\in R$.
\label{linearonreverse}
\end{Lemma}
 \textit{Proof:} The proof is given in the  subsection \ref{proofs} of the Appendix.

Using similar approach, for the higher order, one can observe the following Remark on the linearity of the reverse constraint.
\begin{Remark}
For any positive integer $k$ and $1 \leq i \leq k$, if $\textbf{x}_i \in R^n$, then $\phi^{-1}(\phi(\sum_{i=1}^ka_i\textbf{x}_i)^r) = \sum_{i=1}^ka_i\phi^{-1}(\phi(\textbf{x}_i)^r)$, where $a_i \ \in R$.
\label{linearonreverse_gen}
\end{Remark}

\begin{Corollary}%[Complement]
For any $\textbf{x}, \textbf{y}  \in R^n, \phi^{-1}(\phi(a\textbf{x}+b\textbf{y})^c) = a\phi^{-1}(\phi(\textbf{x})^c) + b\phi^{-1}(\phi(\textbf{y})^c)$ if $a+b \in \{1, 1+2w,3,3+2w\}$, where $a,b \in R$.
\label{linearoncomplement}
\end{Corollary}
 \textit{Proof:}  The proof is given in the  subsection \ref{proofs} of the Appendix.
 %For any $x \in R$, $\phi^{-1}(\phi(x)^c)$ is unique. Note that $(a+b)(2+2w) = 2+2w$ for $a+b \in \{1, 1+2w,3,3+2w\}$, where $a,b \in R$.  Now the proof follows from the Lemma \ref{linearonreverse}.
 
% \begin{figure*}
% \textit{Proof:}
% For any $\textbf{x}, \textbf{y}  \in R^n$, $\textbf{x}=(x_1\ x_2\dots x_n)$ and $\textbf{y}=(y_1\ y_2\dots y_n)$.
% \begin{equation*}
% \begin{split}
% Consider \ \phi(a\textbf{x}+b\textbf{y})^c = & (\phi(ax_1+by_1)^c\ \phi(ax_2+by_2)^c\dots \phi(ax_n+by_n)^c). \\ Thus \
%     \phi^{-1}(\phi(a\textbf{x}+b\textbf{y})^c) = & (\phi^{-1}(\phi(ax_1+by_1)^c)\ \phi^{-1}(\phi(ax_2+by_2)^c)\dots \phi^{-1}(\phi(ax_n+by_n)^c)) \\
%      = & (ax_1+by_1+2+2w\ ax_2+by_2+2+2w\dots ax_n+by_n+2+2w) \\
%      = & (a(x_1+2+2w)+b(y_1+2+2w)\ a(x_2+2+2w)+b(y_2+2+2w)\dots \\
%         & a(x_n+2+2w)+b(y_n+2+2w)) \\
%      = & (a\phi^{-1}(\phi(x_1)^c)+b\phi^{-1}(\phi(y_1)^c)\ a\phi^{-1}(\phi(x_2)^c)+b\phi^{-1}(\phi(y_2)^c)\dots \\
%         & \ldots a\phi^{-1}(\phi(x_n)^c)+b\phi^{-1}(\phi(y_n)^c)) \\
%      = & a((\phi^{-1}(\phi(x_n)^c)\ \phi^{-1}(\phi(x_{n-1})^c)\ \dots \phi^{-1}(\phi(x_1)^c)) \\ 
%         & + b(\phi^{-1}(\phi(y_n)^c)\ \phi^{-1}(\phi(y_{n-1})^c)\dots\phi^{-1}(\phi(y_1)^c)) \\
%      = & a\phi^{-1}(\phi(\textbf{x})^c) + b\phi^{-1}(\phi(\textbf{y})^c).
% \end{split}
% \end{equation*}
% \end{figure*}

Using similar approach, for the higher order, one can observe the following Remark on the linearity for the complement constraint.
\begin{Remark}
For any positive integer $k$ and $1 \leq i \leq k$, if $\textbf{x}_i \in R^n$, then $\phi^{-1}(\phi(\sum_{i=1}^ka_i\textbf{x}_i)^c) = \sum_{i=1}^ka_i\phi^{-1}(\phi(\textbf{x}_i)^c)$ if $\sum_{i=1}^ka_i \in \{1, 1+2w,3,3+2w\}$,  where $a_i \ \in R$.
\label{linearoncomplement_gen}
\end{Remark}

%\begin{Lemma}
%If $\textbf{x}, \textbf{y}, \textbf{2+2w}  \in R^n$ then $\phi^{-1}(\phi(a\textbf{x}+b\textbf{y}))^c = a\phi^{-1}(\phi(\textbf{x}))^c + b\phi^{-1}(\phi(\textbf{y}))^c$, where $a,b\in R$ and $\textbf{2+2w}$ is the string with length $n$ having each entry $2+2w$.
%\end{Lemma}

\begin{Lemma}%[generator matrix reveres]if and if
For any row $\textbf{x} \in G$ over the ring $R$, the DNA code $\phi(<G>_R)$ is closed under reverse if and only if $\phi^{-1}(\phi(\textbf{x})^r) \in$ $<G>_R$, the row span of $G$ over $R$. %where $<G>_R$ is a row span of the matrix $G$ on the ring $R$.
\label{closurereverse}
\end{Lemma}
% \begin{Lemma}%[generator matrix reveres]
% For a matrix $G_{k \times n}$ over the ring $R$ and for any row $\textbf{x} \in G_{k \times n}$ if $\phi^{-1}(\phi(\textbf{x})^r) \in$ $<G>_R$, the row span of $G$ over $R$, then $\phi(<G>_R)$ is closed under reverse. %where $<G>_R$ is a row span of the matrix $G$ on the ring $R$.  
% \label{closurereverse}
% \end{Lemma}

\textit{Proof:}
%The proof follows from the Lemma \ref{linearonreverse}.
\textit{Let $\phi(\textbf{y}) \in \phi(<G>_R)$ = $\phi(\mathscr{C})$ for any $\textbf{y} \in \mathscr{C}$. Thus $\textbf{y} = \sum_{i=1}^k a_i\textbf{x}_i$ for rows $\textbf{x}_i \in G$. Consider $\phi^{-1}(\phi(\textbf{y})^r)$ = $\phi^{-1}(\phi(\sum_{i=1}^k a_i\textbf{x}_i)^r)$ = $\sum_{i=1}^k a_i \phi^{-1}(\phi(\textbf{x}_i)^r) \\ \in \mathscr{C}$ (by using the Remark \ref{linearonreverse_gen}). Thus $\phi^{-1}(\phi(\textbf{y})^r) \in \mathscr{C}$ which directs $\phi(\textbf{y})^r \in \phi(\mathscr{C})$. Hence the DNA code is closed under reverse (by the Remark \ref{closedrev})}. The invert statement is obvious.

\begin{Lemma}%[generator matrix complement]
For a matrix $G$ over the ring $R$, the DNA code $\phi(<G>_R)$ is closed under complement if and only if $\textbf{2+2w}\in <G>_R$, where $\textbf{2+2w}$ is a string with each element $2+2w$.  
\label{closurereversecomp}
\end{Lemma}

\textit{Proof:}
Let $\phi(\textbf{x}) \in \phi(<G>_R) = \phi(\mathscr{C})$ for any $\textbf{x} \in \mathscr{C}$ and $\textbf{2+2w} = (2+2w\ 2+2w\ 2+2w\ \ldots \ 2+2w) \in \mathscr{C}$. Thus $\textbf{x} + \textbf{2+2w} \in \mathscr{C}$ but $\phi^{-1}(\phi(\textbf{x})^c) = \textbf{x} + \textbf{2+2w}$ (by using the Table \ref{gaumapping}). Thus $\phi^{-1}(\phi(\textbf{x})^c) \in \mathscr{C}$ and $\phi(\textbf{x})^c \in \phi(\mathscr{C})$. Hence the DNA code $\mathscr{C}_{DNA}$ is closed under complement using the Remark \ref{closedcomp}.
For inverse statement, if $\phi(\mathscr{C})$ is closed under complement, then $\phi^{-1}(\phi(\textbf{0})^c) \in \mathscr{C}$ (Since $\textbf{0} \in \mathscr{C}$). But $\phi^{-1}(\phi(\textbf{0})^c) = \textbf{0} + \textbf{2+2w}$ = $\textbf{2+2w}$. Hence $\textbf{2+2w} \in \mathscr{C}$. 

The following Theorem is obvious using the Lemma \ref{linearonreverse}, \ref{closurereverse} and \ref{closurereversecomp}.

\begin{Theorem}
Let $\mathscr{C}(n,M,d_{Gau})$ be a code over the ring $R$ with the length $n$, the number of the codewords $M$ and the minimum Gau distance $d_{Gau}$, such that the rows of the generator matrix of $\mathscr{C}$ satisfies the conditions given in the Lemma \ref{closurereverse} and \ref{closurereversecomp}, then $\phi(\mathscr{C})$ is a $\mathscr{C}_{DNA}(2n, M, d_H)$ DNA code with the length $2n$, the number of the codewords $M$ and the minimum Hamming distance $d_{H}$. The DNA code $\mathscr{C}_{DNA}$ satisfies the reverse and reverse complement constraints.
\label{distancepreserving}
\end{Theorem}

% \begin{Theorem}
% Let $\mathscr{C}:(n,M,d_{Gau})$ be a code over the ring $R$ = $\mathbb{Z}_4+w\mathbb{Z}_4$ with $w^2=2+2w$ and $\phi$: $R\rightarrow \Sigma_{DNA}^2$ be any bijective mapping. The collection $\mathscr{C}_{DNA}$ = $\{\phi(\textbf{x}):\forall\ \textbf{x}\in\mathscr{C}\}$ be a DNA code $\mathscr{C}_{DNA}:(n,M,d_{H})$ with parameters $n$ = $2n$, $M$ = $M$ and $d_{H}=d_{Gau}$, where minimum Hamming distance for $\mathscr{C}_{DNA}$ is $d_{H}$ and $\phi(x_i)\in \Sigma_{DNA}^2$ with $i$= $1,2,\ldots,n$. For $\textbf{x}$ = $x_1\ x_2\ldots x_{n-1}\ x_n$, $\phi(\textbf{x})$ = $\phi(x_1)\ \phi(x_2)\ldots\phi(x_{n-1})\ \phi(x_n)$ be a DNA string with length $2n$.
% \end{Theorem}

% \textit{Proof:}
% \textit{For $x \in R$ is mapped to an ordered pair of DNA alphabets through the bijective mapping $\phi:R\rightarrow \Sigma_{DNA}^2$ and $\mathscr{C}_{DNA}$ = $\{\phi(\textbf{x}):\forall\ \textbf{x}\in\mathscr{C}\}$ hence DNA code has length $2n$. One can check that as $\phi$ is a bijective map from $R$ to $\Sigma_{DNA}^2$ implies that $\mathscr{C}_{DNA}$ has M codewords. Moreover, from the Theorem \ref{distanceisometry}, $\phi$ is the distance preserving from $R$ to $\mathscr{C}_{DNA}$ leads to $\mathscr{C}_{DNA}$ has the minimum Hamming distance $d_{H}=d_{Gau}$.}

In the next Section, we construct some new families of the DNA codes.

%%%%%%%%%%%%%%%%%%%%%%%%%%%%%%%%%%%%%%%%%%%%%%%%%%%%%%%%%%%%%%%%%%%%%%%%%%%%%%%%%%%%%%%%%%%%%%%%%%%%%%%%%%%%%%%%%%%%%%%%%%%%%%%%%
\section{Families of DNA Codes from the Ring $R=\mathbb{Z}_4 +w\mathbb{Z}_4$}
By using the results discussed in the Section \ref{codesoverR}, we give new classes of the DNA codes that satisfies the Hamming, reverse and reverse complement constraints. 

\subsection{DNA Codes from the Octacodes Type Codes}
There has been an interesting history of connecting two non-linear binary codes (Kerdock and Preprata) with linear codes over $\mathbb{Z}_4$ \cite{forney1992nordstrom}. Octacode (a linear self dual code of length $n=8$, code size $M=256$ and minimum Lee weight $8$ over $\mathbb{Z}_4$) turns out to be a special case connecting the binary non-linear codes (the Nordstrom-Robinson code) \cite{forney1992nordstrom}. In this section, we construct a code similar to the original Octacode from the ring $\mathbb{Z}_4+w\mathbb{Z}_4$. The self dual code is generated by a generator matrix consisting of the cyclic shifts of the vector $(0 $ $ 2 $ $ 2w $ $ 2+2w $ $  0 $ $ 2 $ $ 2w$ $ 2+2w )$ over $R$. The DNA code is generated from the generator matrix $\mathscr{O}$ of Octacode in the Example \ref{octaexample}. %(see the Table \ref{octadnacode} in Appendix). 
It satisfies the reverse and reverse complement constraints.

\begin{Example}
DNA code $\mathscr{C}_{DNA}(n = 16, M = 64, d_{H} = 8)$ can be obtained from cyclic shifts of the vector $(0 $ $ 2w $ $  2 $ $ 2+2w $ $ 0 $ $ 2w $ $ 2 $ $ 2+2w )$. 
%\end{example}
% \begin{example}
% DNA code $\mathscr{C}_{DNA}$: $ (n,M,d_{H})= (16,64,8)$ is obtained by generator matrix  
\begin{figure*}
  \begin{equation}
\begin{tabular}{cc}
$\mathscr{O} = \left( \begin{array}{cccccccc}
0 & 2w & 2 & 2+2w & 0 & 2w & 2 & 2+2w   \\ 
2+2w & 0 & 2w & 2 & 2+2w & 0 & 2w & 2 \\
2 & 2+2w & 0 & 2w & 2 & 2+2w & 0 & 2w \\
2w & 2 & 2+2w & 0 & 2w & 2 & 2+2w & 0
\end{array} \right)$ &
 \end{tabular}
\end{equation}
\end{figure*}
 \label{octaexample}
\end{Example}

\begin{Remark}
On similar lines, one can obtain Octacodes type DNA codes with parameters listed in the Table \ref{octadnacodeegs} using the different first row vector.
\end{Remark}

\begin{table}[ht]
\centering
\begin{tabular}{|c|c|}
\hline
First Row Vector & DNA Code $\mathscr{C}_{DNA}(n,M,d_H)$	\\ \hline
$(0\ 2\ 2w\ 2+2w)$	&	$(8,16,4)$	\\
$(0\ 2w\  2\  2+2w)$	&	$(8,64,4)$ \\
$(0\  2w\  2\  2+2w\  0\  2w\  2\  2+2w)$	&	$(16,64,8)$ \\
$(0\  2\  2w\   2+2w\   0\  2\  2w\  2+2w)$	&	$(16,16,8)$	\\
\hline
\end{tabular}
\caption{Octacodes types DNA code $\mathscr{C}_{DNA}$.} %$*$ greater value of $M$ (in \cite{mostafanasab2016cyclic} $M=28$).}
\label{octadnacodeegs}
\end{table}
% \vspace{-0.6cm}

DNA codewords generated from the Octacodes have an interesting properties from the application point of view. By removing the trivial codewords (codewords with no GCs or all GCs), one can obtain the DNA codewords with 50$\%$ $GC$ content which are important for the DNA computation.

\subsection{DNA Codes from Simplex Type Codes}
%Simplex codes using rings were introduced in \cite{tapia2008simplex} fisher1941theory,bonnecaze1999cyclic.cengellenmis2010simplex.  %Senary simplex codes was discussed in \cite{gupta2001senary}.  In \cite{cengellenmis2010simplex}, simplex code of type $\gamma$ was introduced. DNA codes satisfying the $GC$ content constraint were studied using simplex codes in \cite{guenda2013construction}. 
Binary simplex codes have a unique geometrical significance (being the dual of the Hamming codes and each having a fixed weight) and were known in $1945$ \cite{fisher1943system} in statistical connections before the actual discovery of the Hamming codes by R.Hamming in $1948$. Simplex codes over rings have been considered by many researchers \cite{gupta1999some}. The simplex codes of type $\alpha$ and $\beta$ over the ring $\mathbb{Z}_4$ have been studied in \cite{bhandari19994}. The simplex codes over the ring $\mathbb{F}_2+ u\mathbb{F}_2$ was given in \cite{al2005simplex}.  %Recently, K. Chatouh et al. generalized the simplex codes over the ring $R_q$ in \cite{chatouh2017simplex}. 
It is natural to study the DNA codes using the simplex codes. DNA codes that avoids the secondary structure formation were designed in \cite{milenkovic2005dna} by using the cyclic simplex codes. In this section, we have designed the DNA codes using the simplex type codes over the ring $\mathbb{Z}_4+w\mathbb{Z}_4$. We give a generator matrix $G^\beta_k$ for the simplex type $\beta$ over the ring $\mathbb{Z}_4+w\mathbb{Z}_4$. The DNA codes which satisfies the reverse and reverse complement constraints is given in the Example \ref{simplexcode_eg}. 

Let $G^\beta_k$ be a matrix over $\mathbb{Z}_4+w\mathbb{Z}_4$ defined inductively by 
 \begin{equation}\label{2}
G^\beta_k = \left( \begin{array}{c|c|c|c}
 0\dots 0       & 2\ldots 2      & 2w\ldots 2w    &2+2w\ldots 2+2w \\ \hline
 G^\beta_{k-1} & G^\beta_{k-1} & G^\beta_{k-1} & G^\beta_{k-1}
\end{array} \right), k \geq 3
\end{equation} with 
\begin{equation}
G^\beta_2 = \left( \begin{array}{cccccccc}
1 & 1 & 1 & 1 & 0 & 2 & 2w & 2+2w   \\ 
0 & 2 & 2w &2+2w  & 1 & 1 & 1 & 1 \end{array} \right).
\end{equation}

Let $S^\beta_k$ be a code generated by the generator matrix of type $\beta$ simplex type code over the ring $R$ = $\mathbb{Z}_4+w\mathbb{Z}_4$ then for $k>1$, $n= 2^{2k-1}$, $M=2^{2k+4}$ and $d_{Gau} = 2^{2k-1}$.

%\begin{Remark}
%If $A_{k-1}$ denotes an array of codewords in $S^\beta_{k-1}$ and if $\textbf{i} = (i\ i\ \ldots \ i)$, where, $i \in \{2,2w,2+2w\}$ then an array of all codewords of $S^\beta_k$ is given by
%\begin{equation}
%    \left( \begin{array}{cccc}
%A_{k-1} & A_{k-1} & A_{k-1} & A_{k-1} \\
%A_{k-1} & \textbf{2} + A_{k-1} & \textbf{2w}+A_{k-1} & \textbf{2+2w}+A_{k-1} \\
%A_{k-1} & \textbf{2w} + A_{k-1} & A_{k-1} & \textbf{2w} + A_{k-1} \\
%A_{k-1} & \textbf{2+2w} + A_{k-1} & \textbf{2w} + A_{k-1} & \textbf{2} + A_{k-1} \\
%\end{array} \right)
%\end{equation}
%\label{codewords_array}
%\end{Remark}

\begin{Theorem}
If $S^\beta_k(n,M,d_{Gau})$ is a simplex $\beta$ type code over the ring $R$ = $\mathbb{Z}_4+w\mathbb{Z}_4$ then the parameters of the corresponding DNA code $\mathscr{C}_{DNA}(n,M,d_H)$ are $n= 2^{2k}$, $M=2^{2k+4}$ and $d_H = 2^{2k-1}$. $\mathscr{C}_{DNA}(n,M,d_H)$ satisfies the reverse and reverse complement constraints.
\label{simplexcodes_theorem}
\end{Theorem}

 \begin{Example}\label{simplexcode_eg}
From the $G^\beta_2$, the DNA codes with parameters $(16,256,8)$ satisfying the reverse and reverse complement constraints can be generated %(See the Table \ref{simplexDNAcodes} in the appendix).
Inductively from $G^\beta_3$, DNA codes with length $n = 64$, $M = 1024$, $d_{H}=32$ is constructed. %(shown in Table \ref{DNAsimplexcode_eg}.
\end{Example}

The DNA codes obtained from the simplex class has purine rich DNA codewords (composition of AGs in the DNA codeword). These DNA codewords are important for the DNA motifs (small length of functional DNA codewords). They also play significant role in the transcription of genes.

\subsection{DNA codes from Reed Muller Type Codes}
The Reed Muller codes (RM) are one of the best known oldest error correcting codes discovered by Reed and Muller in 1954 \cite{muller1954application}. First, the binary Reed Muller codes were introduced and then it was generalized to any $q-$ary alphabets \cite{kasami1968new}. 

The Reed Muller code is denoted as $\mathcal{R}(r,m)$, where $r$ is the order of the code, and $m$ determines the length of code, $n = 2^m$. In this work, we defined the Reed Muller Type codes over the ring $R = \mathbb{Z}_4+w\mathbb{Z}_4$. Using a special type of Reed Muller codes, we give the DNA codes satisfying the reverse and reverse complement constraints. \\
For each positive integers $r$ and $m$ ($1 = r\leq m$), the first order Reed Muller Type code $\mathcal{R}(1,m)$ over $\mathbb{Z}_4+w\mathbb{Z}_4$ where, $w^2$ = $2+2w$ $ \forall $ $ m \geq 1$, is defined by the generator matrix:

\[
G_{1,m+1} =
\begin{pmatrix} 
G_{1,m} & G_{1,m}    \\ 
0 \ldots 0 & z \ldots z \\
\end{pmatrix}, \ m\geq1
\]
\hspace{0.25cm} 
with
 \[
 G_{1,1} =
\begin{pmatrix}
1 & 1    \\ 
0 & z  \\
\end{pmatrix}
\]
where $z \in \{2,w,2+w,2w,2+2w,3w,2+3w\}$.

\begin{Theorem}\label{rmfirstorder}
For the first order Reed Muller Type code $\mathcal{R}(1,m)$ over $\mathbb{Z}_4+w\mathbb{Z}_4$, $\exists$ a DNA code $\mathscr{C}_{DNA}(n,M,d_{H})$ that satisfies reverse and reverse complement constraints with $n$ = $2^{m+1}$, %$M = 16 \times \vert <Z> \vert^m $
\[
M = \left \{ \,
\begin{array}{lll}
2^{m+4} & if & z \in \{2w\},\\
2^{2m+4} & if & z \in \{2,2+2w\}, \\
2^{3m+4} & if & z\in \{w,2+w,3w,2+3w\}, \\
\end{array}
\right.\\
\] and

\[
d_{H} = \left \{ \,
\begin{array}{lll}
2^{m} & if & z \in \{2w,2,2+2w\},\\
2^{m-1} & if & z \in \{w,2+w,3w,2+3w\}. \\
\end{array}
\right. \\
\] %where $z \in \{2,w,2+w,2w,2+2w,3w,2+3w\}$.
\end{Theorem}

\begin{Example}
DNA code $\mathscr{C}_{DNA}(n=16, M=8192, d_{H}=4)$ obtained by the generator matrix 
\begin{equation}
\begin{tabular}{cc}
$G_{1,3} = \left( \begin{array}{cccccccc}
1 & 1 & 1 & 1 & 1 & 1 & 1 & 1   \\ 
0 & 0 & 0 & 0 & w & w & w & w \\
0 & 0 & w & w & 0 & 0 & w & w \\
0 & w & 0 & w & 0 & w & 0 & w 
\end{array} \right)$ &
 \end{tabular} 
 \end{equation}
 \label{reedexample}
 of Reed Muller Type code $\mathcal{R}(1,3)$ satisfies the reverse and reverse complement constraints.
\end{Example}

One can obtain various DNA code of type Reed Muller using different $m$ and $z$ as shown in the Table \ref{reedmullerdnacodeegs}.
\begin{table}[ht]
\centering
\begin{tabular}{|c|c|c|}
\hline
$m$ & Zero divisor $z$ & DNA Code $\mathscr{C}_{DNA}(n,M,d_H)$	\\ \hline
1 & 2 &	$(4,64,2)$	\\
2 & 2 &	$(8,256,4)$ \\
3 & 2 & 	$(16,1024,8)$ \\
2 & $w$ &	$(8,1024,2)$	\\
3 & $w$ & $(16,8192,4)$ \\
\hline
\end{tabular}
\caption{DNA code $\mathscr{C}_{DNA}$ generated by Reed Muller Type code for the zero divisor $z$ with the different values of $m$.} %$*$ greater value of $M$ (in \cite{mostafanasab2016cyclic} $M=28$).}
\label{reedmullerdnacodeegs}
\end{table}

%For Reed Muller codes in Example \ref{reedexample1}, the distance distribution is  $D_{Gau}(\gamma) = \gamma^1 + 5.2\gamma^8 + 52.83 \gamma^{12} + 6\gamma^{16}$.

\subsection{Results}
The DNA codes designed in this work, we have improved the lower bound on $A^{RC,GC}$ and obtained the better code size for some instances compare to DNA codes constructed using the similar rings in the literature.
For $n=8$, $d=4$ and $u=4$, the lower bound obtained for $A^{RC,GC}_4(8,4)$  is \textbf{224} which is greater than the lower bound observed in \cite{chee2008improved,quantumdnacodes} as 128. Hence it is the best known lower bound on $A^{RC,GC}$ obtained for $n=8,d_H=4,u=4$ known in the literature. Results obtained, for the DNA codes, in this work are better than the examples described in \cite{liangcyclic,dinhcyclic,quantumdnacodes}. For an instance the DNA code $\mathscr{C}_{DNA}(n=16,M=8192,d_H=4)$ is better than $\mathscr{C}_{DNA}(n=16,M=28,d_H=4)$ \cite{mostafanasab2016cyclic}. Also the DNA code $\mathscr{C}_{DNA}(n=8,M=256,d_H=4)$ is better than $\mathscr{C}_{DNA}(n=8, M=16, d_H=4)$ \cite{liangcyclic,quantumdnacodes}. One can observe that the Reed Muller Type codes attains the lower bound on size $M$ for $A^{GC}_4(n,d_H,u)$ and $A^{RC,GC}_4(n,d_H,u)$ \cite{king2003bounds} on the DNA code for some values of $n$, $d_H$ and $u = n/2$. For $z = 2$ and $z=2w$, the first order Reed Muller Type code for $(n,d_H,u) = (4,2,2)$ is an optimal code with respect to  $A^{GC}_4(n,d_H,u)$ and $A^{RC,GC}_4(n,d_H,u)$ respectively.

\section{The $r^{th}$ order Reed Muller Type Code}
In order to study the $r^{th}$ order Reed Muller type code, we need the following result.
\begin{Lemma}\label{revmatrix}
Let $G$ and $H$ be two matrices over the ring $R$ such that both the DNA codes $\phi(<G>_R)$ and $\phi(<H>_R)$ are closed under reverse constraint. If each row of $H$ is a row of $G$ then the DNA code $\phi(<T>_R)$ will be closed under reverse, where
\[
T =
\begin{pmatrix}
G & G     \\ 
\textbf{0} & H \\
\end{pmatrix}
\]
\end{Lemma}

% \textit{Proof:} Any row of $T$ is the row of either matrix ($G \ G$) or the matrix (\textbf{0} \ $H$). Now consider two cases for this:\\
% \textit{Case: 1} If $(\textbf{x} \ \textbf{x})$ is the row of the matrix ($G \ G$) then $\textbf{x}$ will be the row of the matrix $G$. Using the Lemma \ref{closurereverse}, $\phi^{-1}(\phi(\textbf{x})^r) \in <G>_R$ and therefore ($\phi^{-1}(\phi(\textbf{x})^r)$ \ $\phi^{-1}(\phi(\textbf{x})^r)$) $\in <G \ G>_R$ by using the Lemma \ref{concate_rev} for each row of the matrix $G$. \\
% \textit{Case: 2} If $(\textbf{0} \  \textbf{y})$ is the row of the matrix (\textbf{0} $H$) then $\textbf{y}$ will be the row of $H$. By using the Lemma \ref{closurereverse}, $\phi^{-1}(\phi(\textbf{y})^r) \in <H>_R$. Therefore $(\textbf{0} \ \phi^{-1}(\phi(\textbf{x})^r)) \in \ <\textbf{0} \ H>_R  \subseteq <T>_R$. But $\textbf{y}$ is also the row of $G$ so from the case 1, ($\phi^{-1}(\phi(\textbf{y})^r)$ \ $\phi^{-1}(\phi(\textbf{y})^r)$) $\in <G \ G>_R \subseteq <T>_R$. Thus ($\phi^{-1}(\phi(\textbf{y})^r)$ \ \textbf{0}) $\in <T>_R$ for each row \textbf{y} of the matrix $H$. Now by case 1 and case 2, it is concluded that $\phi^{-1}(\phi(\textbf{t})^r) \in <T>$ for each row $\textbf{t}$ of the matrix $T$. Hence by using the Lemma \ref{closurereverse}, the DNA code $\phi(<T>_R)$ is closed under reverse.

One can also define $r^{th}$ order Reed Muller type code as follows.
For the given zero divisor $z\in R$ and each integers $r$, $m$ ($0 \leq r \leq m$), the $r^{th}$ order Reed Muller code $\mathcal{R}(r,m)$ over the ring $R$ is defined by the generator matrix
\[
G_{r,m} =
\begin{pmatrix}
G_{r,m-1} & G_{r,m-1}    \\ 
0 & G_{r-1,m-1} \\
\end{pmatrix},\ 1 \leq r \leq m-1.
\]
with
\[
 G_{m,m}\ = 
 \left(
 \begin{array}{c}
      G_{m-1,m}  \\
      0\ 0\ldots\ 0\ z
 \end{array}
 \right)
 \]
and $G_{0,m}\ = \left(1\ 1\ldots 1\right)$ is the all one matrix with length $2^m$.
 
\begin{Theorem}\label{gen_r_rm}
For the $r^{th}$ order Reed Muller code $\mathcal{R}(r,m)$ over the ring $R$, the DNA code $\mathscr{C}_{DNA}=\phi(\mathcal{R}(r,m))$ satisfies the reverse and reverse complement constraints.
The $(n,M,d_H)$ parameters of the DNA code $\mathscr{C}_{DNA}$ are $n$ = $2^{m+1}$, %$M = 16 \times \vert <Z> \vert^m $
\[
M = \left \{ \,
\begin{array}{lll}
2^{4b-3a} & if & z \in \{2w\},\\
2^{4b-2a} & if & z \in \{2,2+2w\}, \\
2^{4b-a} & if & z\in \{w,2+w,3w,2+3w\}, \\
\end{array}
\right.\\
\] and

\[
d_{H} = \left \{ \,
\begin{array}{lll}
2^{m-r+1} & if & z \in \{2w,2,2+2w\},\\
2^{m-r} & if & z \in \{w,2+w,3w,2+3w\}, \\
\end{array}
\right. \\
\] where $a = \sum_{i=0}^{r-1}\binom{m-1}{i}$ and $b = \sum_{i=0}^r\binom{m}{i}$.
\end{Theorem}

The results obtained on the reverse and reverse complement constraints for the family of Octa type and Simplex type DNA codes suggest the following general Theorems and Remarks.

For a positive integer $k$, let $P$ be a matrix over the ring $R$ with $4k$ length vector $(2\ 2\ldots2) \in \ <P>_R$. For $i=1,2,3,4$, all the four elements $z_i\in\{0,2,2w,2+2w\}$ are distinct and all the four vectors \textbf{z}$_i=(z_i\ z_i\ldots z_i)$ have same length $k$. Now, consider the matrix 
\begin{equation*}
\begin{tabular}{cc}
$G = \left( 
\begin{array}{c}
\begin{array}{cccc}
\textbf{z}_1 & \textbf{z}_2 & \textbf{z}_3 & \textbf{z}_4   
\end{array}  \\
P
\end{array}
\right)$. &
 \end{tabular} 
 \end{equation*}
 For the matrix $G$, parameters and constraints are discussed in the following Theorems \ref{gen_para} and \ref{genconstraints}.  
 
\begin{Theorem}\label{gen_para} %(\textbf{New added})
 If the parameters of the DNA code $\phi(<P>_R)$ are ($8k,\ M^P,\ d_H^P$) then the parameters of the DNA code $\phi(<G>_R)$ are ($8k,\ M^G,\ d_H^G$), where $d_H^G\leq min\{4k,\ d^P\}$ and 
\[
M^G = \left\{ \begin{array}{ll}
M^P &  \mbox{ if } (\textbf{z}_1\ \textbf{z}_2\ \textbf{z}_3\ \textbf{z}_4)\in<P>_R, \\
4M^P & \mbox{ if } (\textbf{z}_1\ \textbf{z}_2\ \textbf{z}_3\ \textbf{z}_4)\notin<P>_R. \\
\end{array}\right.
\]
\end{Theorem}
% \textit{Proof:}
% Both the matrices $G$ and $P$ have $4k$ number of columns so, by using the Theorem \ref{distancepreserving}, the length of a codeword of the DNA code $\phi(<G>_R)$ is $8k$. Let the matrix $P$ be of type $\{k_0, k_1,k_2,k_3\}$. If $(\textbf{z}_1\ \textbf{z}_2\ \textbf{z}_3\ \textbf{z}_4)\in<P>_R$ then the matrix $G$ is of type $\{k_0, k_1,k_2,k_3\}$ and if $(\textbf{z}_1\ \textbf{z}_2\ \textbf{z}_3\ \textbf{z}_4)\notin<P>_R$ then the matrix $G$ is of type $\{k_0, k_1,k_2+1,k_3\}$. Hence, the result holds for the code size $M$. The minimum Gau distance for $<\textbf{z}_1\ \textbf{z}_2\ \textbf{z}_3\ \textbf{z}_4>_R$, is $4k$ and the minimum Gau distance for $<P>_R$, is $d^P_{Gau}$. Hence, the minimum Gau distance for $<G>_R$ is bounded by $min\{4k,\ d^P\}$. Using the Theorem \ref{distancepreserving}, the result holds.

\begin{Theorem} %(\textbf{New added})
If the DNA code $\phi(<P>_R)$ is closed under the reverse constraint then the DNA code $\phi(<G>_R)$ will be closed under the reverse and reverse complement constraints.
\label{genconstraints}
\end{Theorem}

\begin{Remark} %\textbf{New Added}
In the Theorem \ref{gen_para}, one can obtain the DNA code with the higher parameters using induction on the matrix $G$ with the base case $P$ such that the DNA code satisfies reverse and reverse complement constraints.
\end{Remark}

\begin{Lemma}\label{concate_rev} %(\textbf{New added})
For a matrix $G_1$ over the ring $R$, if the DNA code $\phi(<G_1>_R)$ is closed under reverse constraint then the DNA code $\phi(<G_k>_R)$ will be closed under reverse constraint, where $G_k = (G _{k-1}\ G_1)$ for $k > 1$.
\end{Lemma}
% \textit{Proof:}
% For each \textbf{x}$\in <G_1>_R$, $\phi^{-1}(\phi($\textbf{x}$)^r)\in <G_1>_R$.
% Note that, the matrix $G_k$ is $k$ times block repetition of the matrix $G_1$, so (\textbf{x} \textbf{x} $\ldots$ \textbf{x})$\in G_k$ and ($\phi^{-1}(\phi($\textbf{x}$)^r)$ $\phi^{-1}(\phi($\textbf{x}$)^r)$ $\ldots$ $\phi^{-1}(\phi($\textbf{x})$)^r)\in G_k$, for every \textbf{x}$\in G_1$. 
% But, ($\phi^{-1}(\phi($\textbf{x}$)^r)$ $\phi^{-1}(\phi($\textbf{x}$)^r)$ $\ldots$ $\phi^{-1}(\phi($\textbf{x})$)^r)$ = $\phi^{-1}(\phi($\textbf{x} \textbf{x} $\ldots$ \textbf{x}$)^r)$ which directs $\phi^{-1}(\phi($\textbf{x} \textbf{x} $\ldots$ \textbf{x}$)^r)\in<G_k>_R$. Using the Remark \ref{closedrev}, the result holds.

% The matrix $G_1$ is closed under the reverse constraint, so $\phi^{-1}(\phi($\textbf{x}$)^r)\in <G_1>_R$, for each \textbf{x}$\in <G_1>_R$. Since the matrix $G_k$ is $k$ times block repetition of the matrix $G_1$, (\textbf{x} \textbf{x} $\ldots$ \textbf{x})$\in G_k$ and ($\phi^{-1}(\phi($\textbf{x}$)^r)$ $\phi^{-1}(\phi($\textbf{x}$)^r)$ $\ldots$ $\phi^{-1}(\phi($\textbf{x})$)^r)\in G_k$, for each \textbf{x}$\in G_1$. Note that ($\phi^{-1}(\phi($\textbf{x}$)^r)$ $\phi^{-1}(\phi($\textbf{x}$)^r)$ $\ldots$ $\phi^{-1}(\phi($\textbf{x})$)^r)$ = $\phi^{-1}(\phi($\textbf{x} \textbf{x} $\ldots$ \textbf{x}$)^r)$.

\begin{Remark}\label{comple_induc} % (\textbf{New added}) 
For a matrix $G_1$ over the ring $R$, if \textbf{2+2w} = $(2+2w\ 2+2w\ldots 2+2w)\in <G_1>_R$ then (\textbf{2+2w} \textbf{2+2w}...\textbf{2+2w})$\in <G_k>_R$, where $G_k = (G _{k-1}\ G_1)$ for $k > 1$.
Hence, if the DNA code $\phi(<G_1>_R)$ is closed under complement constraint then the DNA code $\phi(<G_k>_R)$ will be closed under complement constraint.
\end{Remark}

\section{Conclusion}
This paper scratches an interesting area of the DNA codes using the ring $R$ = $\mathbb{Z}_4 + w\mathbb{Z}_4$, where $w^2=2+2w$. A new distance called the Gau distance on the ring $R$ is introduced. We have also proposed a new distance preserving Gau map $\phi$ from the elements of the ring $R$ to all the DNA codewords of length 2. Several new families of the DNA codes are obtained. Some of them are optimal with respect to the bounds and are better than the DNA codes obtained in the literature. For the future study, it would be an interesting task to investigate the algebraic structure of the cyclic codes over the ring $R$ and their correspondence to the DNA codes using the map $\phi$. Using algebraic coding, constructing the optimal DNA codes meeting the bounds on reverse, reverse complement, $GC$ content constraints is also an interesting future work.

%\section*{Acknowledgment}
%
%We are indebted to Michael Shell for maintaining and improving
%\texttt{IEEEtran.cls}. 

%%%%%%
%% To balance the columns at the last page of the paper use this
%% command:
%%
%\enlargethispage{-1.2cm} 
%%
%% If the balancing should occur in the middle of the references, use
%% the following trigger:
%%
%\IEEEtriggeratref{3}
%%
%% which triggers a \newpage (i.e., new column) just before the given
%% reference number. Note that you need to adapt this if you modify
%% the paper.  The "triggered" command can be changed if desired:
%%
%\IEEEtriggercmd{\enlargethispage{-20cm}}
%%
%%%%%%

%%%%%%
%% References:
%% We recommend the usage of BibTeX:
%%
%\bibliographystyle{IEEEtran}
%\bibliography{definitions,bibliofile}
%%
%% where we here have assume the existence of the files
%% definitions.bib and bibliofile.bib.
%% BibTeX documentation can be obtained at:
%% http://www.ctan.org/tex-archive/biblio/bibtex/contrib/doc/
%%%%%%

%% Or you use manual references (pay attention to consistency and the
%% formatting style!):

\bibliographystyle{IEEEtran}
\bibliography{dnacodesreferences}

\newpage
 $ $
\newpage
\section{Appendix}
\subsection{Proofs}\label{proofs}

\subsubsection{The proof of the Theorem \ref{distanceisometry}}
%\textit{Proof:}
We prove it for $n=1$. Higher case is obvious. For $x, y \in R$, let $x = m_{i,j} \in \mathscr{M}$, $y = m_{i^{'},j^{'}} \in \mathscr{M}$ for some $0 \leq i,j \leq 3$ and $0 \leq i^{'},j^{'} \leq 3 $. Let us discuss the different cases of $x,y \in R$.
%For $x,y \in R$ and  $\phi(x)$, $\phi(y) \in \Sigma_{DNA}^{2n}$, let $\mathscr{M}(x) = (x_i,x_j)$ and $\mathscr{M}(y) = (y_i,y_j)$ be the indexes of the elements $x$ and $y$ of the matrix $\mathscr{M}$. Let us discuss different case of $x,y \in R $.
\begin{enumerate}
    \item If $i = i^{'}$ and $j = j^{'}$, then $x = y$ and $\phi(x) = \phi(y) \implies$ $d_H(\phi(x), \phi(y)) =0$ and $d_{Gau}(x,y) = 0$.
    \item If $i \neq i^{'}$ and $j = j^{'}$, then $d_H(\phi(x), \phi(y)) = 1$ and $d_{Gau}(x,y) = 1$.
    \item If $i = i^{'}$ and $j \neq j^{'}$, then $d_H(\phi(x), \phi(y)) = 1$ and $d_{Gau}(x,y) = 1$.
    \item If $i \neq i^{'}$ and $j \neq j^{'}$ then $d_H(\phi(x), \phi(y)) = 2$ and $d_{Gau}(x,y) = 2$.
\end{enumerate}
% \begin{enumerate}
%     \item If $x_i = y_i$ and $x_j = y_j$, then $x = y$ and $\phi(x) = \phi(y) \implies$ $d_H(\phi(x), \phi(y)) =0$ and $d_{Gau}(x,y) = 0$.
%     \item If $x_i \neq y_i$ and $x_j = y_j$, then $d_H(\phi(x), \phi(y)) = 1$ and $d_{Gau}(x,y) = 1$.
%     \item If $x_i = y_i$ and $x_j \neq y_j$, then $d_H(\phi(x), \phi(y)) = 1$ and $d_{Gau}(x,y) = 1$.
%     \item If $x_i \neq y_i$ and $x_j \neq y_j$ then $d_H(\phi(x), \phi(y)) = 2$ and $d_{Gau}(x,y) = 2$.
%     %
% \end{enumerate}
Considering all the above cases, it is obvious that $\phi: (R^n, d_{Gau})$ to $(\Sigma_{DNA}^{2n},d_H)$ is an isometry.

\subsubsection{The proof of the Theorem \ref{simplexcodes_theorem}}
%\textit{Proof:}
The proof has two parts. The first part contains the proof for the parameters of the DNA code $\mathscr{C}_{DNA} = \phi(S^{\beta}_k)$. The second part proves that the DNA code satisfies reverse and reverse complement constraints.
\begin{enumerate}
    \item By using the induction on $k$, one can observe that the length of the DNA code $\mathscr{C}_{DNA}$ is $n= 2^{2k}$ with the initial condition on $G^\beta_2$. The matrix $G^\beta_k$ is type $\{2, 0, k-2, 0\}$ matrix. Hence, the number of DNA codewords $M=2^{2k+4}$. %For $\textbf{x} \in G^\beta_2$, $d_{Gau}(z(\textbf{x}), \textbf{0}) = n/2$, where $z \in \{0,2,2w,2+2w\}$. 
     For $G_k^\beta$, the minimum Gau distance is proved by using induction on $k$. For $k=2$, the base case $G_2^\beta$ is trivial. 
    For some positive integer $k-1$, let the minimum Gau distance of $S_{k-1}^\beta$ is $2^{2(k-1)-1}$. For each $z \in \{0,2,2w,2+2w\}$, note that $(z\ z\ldots z)\in S_{k-1}^\beta$. Thus, by the matrix structure, the minimum Gau distance of $ S_{k}^\beta$ can not be more than four times of $(2^{2(k-1)-1})$ ie. $2^{2k-1}$. But, $d_{Gau}($\textbf{0}$,z$\textbf{x}$_{k})=2^{2k-1}$, where \textbf{0} is all zero vector and \textbf{x}$_{k}$ is the last row of the matrix $G_{k}^\beta$. Hence, the minimum Gau distance of $S_{k}^\beta$ is $2^{2k-1}$. Using the Theorem \ref{distancepreserving}, the result holds for the Hamming distance of the DNA code $\mathscr{C}_{DNA}$.
\item To prove the closure of complement, observe that \textbf{2+2w} is 2+2w times the sum of the last two rows of $G^\beta_k$. Hence $\textbf{2+2w} \in S^{\beta}_k$. Thus by the Lemma \ref{closurereversecomp}, the DNA code $\mathscr{C}_{DNA}$ is closed under complement. The closure of $\mathscr{C}_{DNA}$ with respect to the reverse can be proved by using the Lemma \ref{closurereverse}. Let $i^{th}$ row of the matrix $G^\beta_k$ be $\textbf{x}_i$, where $i = 1, 2, \ldots k$. For each $\textbf{x}_i \in G^\beta_k$, observe that
\begin{scriptsize}
\[
\phi^{-1}(\phi(\textbf{x}_i)^r) = \left\{ \begin{array}{ll}
(2+2w)(\textbf{x}_k + \textbf{x}_{k-1}) + \textbf{x}_i &  \mbox{ if }i=1,2, \ldots k-2 , \\
(2+2w)\textbf{x}_{k-1} + 3\textbf{x}_{k} & \mbox{ if }i=k-1, \\
(2+2w)\textbf{x}_k + 3\textbf{x}_{k-1} &   \mbox{ if }i=k. 
\end{array}\right.
\]
\end{scriptsize}
 As $2+2w, \textbf{x}_k, \textbf{x}_{k-1} \in S^{\beta}_k$, $\phi^{-1}(\phi(\textbf{x}_i)^r) \in S^{\beta}_k$. Thus $\mathscr{C}_{DNA}$ satisfy reverse constraint (from the Remark \ref{closurereverse}).  Hence, by using the Remark \ref{r_rc_exist}, the DNA code satisfies reverse and reverse complement constraints.
  \end{enumerate}

\subsubsection{Proof of the Theorem \ref{rmfirstorder}}
%\textit{Proof:}
%For the given generator matrix  $G_{1,m+1}$, by the construction of the matrix $G_{1,m+1}$, the length $n = 2^m \times 2$, hence the length of the code is $n=2^{m+1}$. Number of codewords $M = 16. |<z>|^m$ where, $|<z>|$ is the size of the ideals of the zero divisor $z$.
The proof has two parts. The first part contains the proof for the parameters of the DNA code $\mathscr{C}_{DNA}$ = $\phi(\mathcal{R}(1,m))$. The second part proves that the DNA codes satisfies reverse and reverse complement constraints.

\begin{enumerate}
   \item Using induction on $m$, one can observe that the length $n=2^{m+1}$ of $\mathscr{C}_{DNA}$ with the base case $G_{1,1}$. For $z \in \{w,2+w,3w,2+3w\}$, the matrix $G_{1,m}$ is of type $\{1,m,0,0\}$, for $z \in \{2,2+2w\}$, the matrix $G_{1,m}$ is of type $\{1,0,m,0\}$ and for $z \in \{2w\}$, the matrix $G_{1,m}$ is of type $\{1,0,0,m\}$, hence the result holds for number of codewords $M$ (using the Theorem \ref{distancepreserving}). Due to the symmetry of the matrix $G_{1,m}$, note that any two codewords differ at least at $2^{m-1}$ positions. Hence, the minimum Gau distance $d_{Gau}$ is $d2^{m-1}$, where $d=min\{d_{Gau}(x,y):x\neq y\mbox{ and } x,y\in<z>\}$. % (because observe that $(0\ 0\ldots\ 0)$ and $(0\ 0\ldots\ 0\ z\ z\ldots\ z)$ are in $<G_{1,m}>_R)$. % 2^m \geq d_{Gau} \geq 2^{m-1}$. 
Therefore for all the different cases of zero divisors $z$, the results hold.
\item For each integer $m\geq1$, the codeword \textbf{2+2w} is obtained by multiplying $2+2w$ to the row \textbf{1} of the matrix $G_{1,m}$ over the ring $R$, where \textbf{1} = $(1\ 1\ldots 1)$. From the Lemma \ref{closurereversecomp}, the DNA code $\mathscr{C}_{DNA}$ is closed under complement. The closure of $\mathscr{C}_{DNA}$ with respect to reverse can be proved by using the Lemma \ref{closurereverse}. Let $\textbf{x}_i$ be the $i^{th}$ row of the matrix $G_{1,m}$, then for each  $\textbf{x}_i \in G_{1,m}$, observe that 
\[
\phi^{-1}(\phi(\textbf{x}_i)^r) = \left\{ \begin{array}{ll}
\textbf{3} &  \mbox{ if } i = 1 , \\
3\textbf{z} + \textbf{x}_{i} & \mbox{ if }i=2,3, \ldots m+1. \\
\end{array}\right.
\]
Therefore $\phi^{-1}(\phi(\textbf{x}_i)^r)\in\mathcal{R}(1,m)$ because \textbf{x}$_i$, \textbf{3}, \textbf{z} $\in\mathcal{R}(1,m)$ for each $i=1,2,\ldots,m+1$.
Thus from the Lemma \ref{closurereverse}, the DNA code $\mathscr{C}_{DNA}$ is closed under reverse. Hence, by using the Remark \ref{r_rc_exist}, the DNA code satisfies reverse and reverse complement constraints.
\end{enumerate}

\subsubsection{The proof of the Lemma \ref{revmatrix}}
%\textit{Proof:} 
Any row of $T$ is the row of either matrix ($G \ G$) or the matrix (\textbf{0} \ $H$). Now consider two cases for this:\\
\textit{Case: 1} If $(\textbf{x} \ \textbf{x})$ is the row of the matrix ($G \ G$) then $\textbf{x}$ will be the row of the matrix $G$. Using the Lemma \ref{closurereverse}, $\phi^{-1}(\phi(\textbf{x})^r) \in <G>_R$ and therefore ($\phi^{-1}(\phi(\textbf{x})^r)$ \ $\phi^{-1}(\phi(\textbf{x})^r)$) $\in <G \ G>_R$ by using the Lemma \ref{concate_rev} for each row of the matrix $G$. \\
\textit{Case: 2} If $(\textbf{0} \  \textbf{y})$ is the row of the matrix (\textbf{0} $H$) then $\textbf{y}$ will be the row of $H$. By using the Lemma \ref{closurereverse}, $\phi^{-1}(\phi(\textbf{y})^r) \in <H>_R$. Therefore $(\textbf{0} \ \phi^{-1}(\phi(\textbf{x})^r)) \in \ <\textbf{0} \ H>_R  \subseteq <T>_R$. But $\textbf{y}$ is also the row of $G$ so from the case 1, ($\phi^{-1}(\phi(\textbf{y})^r)$ \ $\phi^{-1}(\phi(\textbf{y})^r)$) $\in <G \ G>_R \subseteq <T>_R$. Thus ($\phi^{-1}(\phi(\textbf{y})^r)$ \ \textbf{0}) $\in <T>_R$ for each row \textbf{y} of the matrix $H$. Now by case 1 and case 2, it is concluded that $\phi^{-1}(\phi(\textbf{t})^r) \in <T>$ for each row $\textbf{t}$ of the matrix $T$. Hence by using the Lemma \ref{closurereverse}, the DNA code $\phi(<T>_R)$ is closed under reverse.

\subsubsection{The proof of the Theorem \ref{gen_r_rm} }
%\textit{Proof:}
The proof of the Theorem follows in two parts. The first part proves the parameters of the DNA code $\mathscr{C}_{DNA}$ and the second part proves the reverse and reverse complement constraints. 
\begin{enumerate}
    \item Using induction on $m$, one can observe that the length $n=2^{m+1}$ of $\mathscr{C}_{DNA}$ with the base case $G_{1,1}$.
    Note that in the matrix $G_{r,m}$, the total number of rows which contain the zero divisor $z$ are $b = \sum_{i=0}^r\binom{m}{i}$ and the total number of rows are $a = \sum_{i=0}^{r-1}\binom{m-1}{i}$.
    For $z \in \{w,2+w,3w,2+3w\}$, the matrix $G_{r,m}$ is of type $\{b-a,a,0,0\}$, for $z \in \{2,2+2w\}$, the matrix $G_{r,m}$ is of type $\{b-a,0,a,0\}$ and for $z \in \{2w\}$, the matrix $G_{r,m}$ is of type $\{b-a,0,0,a\}$, hence the result holds for number of codewords $M$.
    Due to the symmetry of the matrix $G_{r,m}$, note that any two codewords are differ at least at $2^{m-1}$ positions.
    Hence, the minimum Gau distance $d_{Gau}$ is $d2^{m-1}$, where $d=min\{d_{Gau}(x,y):x\neq y\mbox{ and } x,y\in<z>\}$. % (because observe that $(0\ 0\ldots\ 0)$ and $(0\ 0\ldots\ 0\ z\ z\ldots\ z)$ are in $<G_{1,m}>_R)$. % 2^m \geq d_{Gau} \geq 2^{m-1}$. 
Therefore for all the different cases of zero divisor $z$, the result holds for the distance by using the Theorem \ref{distancepreserving}.
    \item  For each integers $r$, $m$ ($0 \leq r \leq m$), the codeword \textbf{2+2w} is obtained by multiplying $2+2w$ to the row \textbf{1} of the matrix $G_{1,m}$ over the ring $R$, where \textbf{1} = $(1\ 1\ldots 1)$. 
     For the DNA code $\mathscr{C}_{DNA}$, the reverse constraint can be proved using induction on $r$. 
    For any integer $m\geq0$, the matrix $G_{0,m}=(1\ 1\ldots1)$ with $2^m$ columns and therefore $(3\ 3\ldots3)\in R(0,m)$. 
    Using the Lemma \ref{closurereverse}, the DNA code $\phi(R(0,m))$ is closed under the reverse constraint. 
    Now, assume that the DNA code $\phi(R(r-1,m))$ is closed under reverse for each integer $m\geq r-1$. 
    For the given integer $r$, the reverse constraint of $\phi(R(r,m))$ can be proved using induction on $m\ (\geq r)$. 
    For the base case $m=r$, we will prove that $\phi(R(r,r))$ is closed under reverse. 
    Note that 
    \[
    \begin{array}{cc}
    G_{r,r}=
    \left(
    \begin{array}{c}
         G_{r-1,r}  \\
         0\ 0\ldots 0 
    \end{array}
    \right) = 
    &
    \left(
    \begin{array}{cc}
        G_{r-1,r-1} & G_{r-1,r-1} \\
        \textbf{0} & G_{r-1,r-1}
    \end{array}
    \right)
    \end{array}
    \]
    Using the Lemma \ref{revmatrix}, $\phi(R(r,r))$ is closed under reverse. 
    By recurrence construction of the matrix, each row of the matrix $G_{r-1,m-1}$ is the row of the matrix $G_{r,m-1}$. 
    Using the Lemma \ref{revmatrix}, the DNA code $\mathscr{C}_{DNA}$ is closed under reverse.
\end{enumerate}

\subsubsection{The proof of the Theorem \ref{gen_para}}
%\textit{Proof:}
Both the matrices $G$ and $P$ have $4k$ number of columns so, by using the Theorem \ref{distancepreserving}, the length of a codeword of the DNA code $\phi(<G>_R)$ is $8k$. Let the matrix $P$ be of type $\{k_0, k_1,k_2,k_3\}$. If $(\textbf{z}_1\ \textbf{z}_2\ \textbf{z}_3\ \textbf{z}_4)\in<P>_R$ then the matrix $G$ is of type $\{k_0, k_1,k_2,k_3\}$ and if $(\textbf{z}_1\ \textbf{z}_2\ \textbf{z}_3\ \textbf{z}_4)\notin<P>_R$ then the matrix $G$ is of type $\{k_0, k_1,k_2+1,k_3\}$. Hence, the result holds for the code size $M$. The minimum Gau distance for $<\textbf{z}_1\ \textbf{z}_2\ \textbf{z}_3\ \textbf{z}_4>_R$, is $4k$ and the minimum Gau distance for $<P>_R$, is $d^P_{Gau}$. Hence, the minimum Gau distance for $<G>_R$ is bounded by $min\{4k,\ d^P\}$. Using the Theorem \ref{distancepreserving}, the result holds.

\subsubsection{The proof of the Theorem \ref{genconstraints}}
%\textit{Proof:}
For the complement constraint, note that $(2+2w\ 2+2w\ldots2+2w)\in<G>_R$ because $(2\ 2\ldots2)\in<P>_R$.
Using the Lemma \ref{closurereversecomp}, the DNA code $\phi(<G>_R)$ is closed under complement constraint.
For the reverse constraint, consider $\phi^{-1}(\phi(\textbf{z}_1\ \textbf{z}_2\ \textbf{z}_3\ \textbf{z}_4)^r)=(\textbf{z}_4\ \textbf{z}_3\ \textbf{z}_2\ \textbf{z}_1)$ because $\phi^{-1}(\phi(z)^r)=z$, for any $z\in\{0,2,2w,2+2w\}$. 
 For the elements $0,2,2w,2+2w$ of the ring $R$, note that the sum of any two elements is equal to the sum of the another two elements. 
If the length of the vectors $\textbf{z}_1,\textbf{z}_2,\textbf{z}_3,\textbf{z}_4$ is same then $\textbf{z}_1+\textbf{z}_4=\textbf{z}_2+\textbf{z}_3$. 
If $\textbf{z}_1+\textbf{z}_4=\textbf{z}_2+\textbf{z}_3=\textbf{z}$ then $(\textbf{z}_1\ \textbf{z}_2\ \textbf{z}_3\ \textbf{z}_4)+(\textbf{z}_4\ \textbf{z}_3\ \textbf{z}_2\ \textbf{z}_1)=(\textbf{z}\ \textbf{z}\ \textbf{z}\ \textbf{z})$ for some $z\in\{2,2w,2+2w\}$, where \textbf{z} = $(z\ z\ldots z)$ is $k$ length codeword.
 %only if all the five vectors $\textbf{z},\textbf{z}_i$ ($i=1,2,3,4$) have same length. 
 But, $(2\ 2\ldots2)\in <P>_R$ so $(\textbf{z}\ \textbf{z}\ \textbf{z}\ \textbf{z})\in <P>_R$ and therefore$(\textbf{z}_4\ \textbf{z}_3\ \textbf{z}_2\ \textbf{z}_1)$ = $(\textbf{z}\ \textbf{z}\ \textbf{z}\ \textbf{z})-(\textbf{z}_1\ \textbf{z}_2\ \textbf{z}_3\ \textbf{z}_4)\in <G>_R$ which directs $\phi^{-1}(\phi(\textbf{z}_1\ \textbf{z}_2\ \textbf{z}_3\ \textbf{z}_4)^r)\ \in \ <G>_R$. 
 By using the Lemma \ref{closurereverse}, the DNA code $\phi(<G>_R)$ is closed under reverse constraint. Thus by using the Remark \ref{r_rc_exist}, the results holds.
 
\subsubsection{The proof of the Lemma \ref{concate_rev}}
%\textit{Proof:}
For each \textbf{x}$\in <G_1>_R$, $\phi^{-1}(\phi($\textbf{x}$)^r)\in <G_1>_R$.
Note that, the matrix $G_k$ is $k$ times block repetition of the matrix $G_1$, so (\textbf{x} \textbf{x} $\ldots$ \textbf{x})$\in G_k$ and ($\phi^{-1}(\phi($\textbf{x}$)^r)$ $\phi^{-1}(\phi($\textbf{x}$)^r)$ $\ldots$ $\phi^{-1}(\phi($\textbf{x})$)^r)\in G_k$, for every \textbf{x}$\in G_1$. 
But, ($\phi^{-1}(\phi($\textbf{x}$)^r)$ $\phi^{-1}(\phi($\textbf{x}$)^r)$ $\ldots$ $\phi^{-1}(\phi($\textbf{x})$)^r)$ = $\phi^{-1}(\phi($\textbf{x} \textbf{x} $\ldots$ \textbf{x}$)^r)$ which directs $\phi^{-1}(\phi($\textbf{x} \textbf{x} $\ldots$ \textbf{x}$)^r)\in<G_k>_R$. Using the Remark \ref{closedrev}, the result holds.

\newpage
\subsubsection{The proof of the Lemma \ref{linearonreverse}}
%\begin{figure*}
%\textit{Proof:}
For any $\textbf{x}, \textbf{y}  \in R^n$, $\textbf{x}=(x_1\ x_2\dots x_n)$ and $\textbf{y}=(y_1\ y_2\dots y_n)$. 
\begin{equation*}
\begin{split}
Consider \ \phi(a\textbf{x}+b\textbf{y})^r = & (\phi(ax_n+by_n)^r\ \phi(ax_{n-1}+by_{n-1})^r\dots \phi(ax_1+by_1)^r). \\ Thus \
    \phi^{-1}(\phi(a\textbf{x}+b\textbf{y})^r) = & (\phi^{-1}(\phi(ax_n+by_n)^r)\ \phi^{-1}(\phi(ax_{n-1}+by_{n-1})^r)\dots \phi^{-1}(\phi(ax_1+by_1)^r)) \\
     = & ((3ax_n+3by_n)\ (3ax_{n-1}+3by_{n-1})\dots (3ax_1+3by_1)) \\
     = & (a\phi^{-1}(\phi(x_n)^r)+b\phi^{-1}(\phi(y_n)^r)\ a\phi^{-1}(\phi(x_{n-1})^r)+b\phi^{-1}(\phi(y_{n-1})^r)\dots \\ 
        & \ldots a\phi^{-1}(\phi(x_1)^r)+b\phi^{-1}(\phi(y_1)^r)) \\
     = & a(\phi^{-1}(\phi(x_n)^r)\ \phi^{-1}(\phi(x_{n-1})^r)\ \dots \phi^{-1}(\phi(x_1)^r)) \\ &+ b(\phi^{-1}(\phi(y_n)^r)\ \phi^{-1}(\phi(y_{n-1})^r)\dots \phi^{-1}(\phi(y_1)^r)) \\
     = & a\phi^{-1}(\phi(\textbf{x})^r) + b\phi^{-1}(\phi(\textbf{y})^r).
\end{split}
\end{equation*}
%\end{figure*}

\subsubsection{The proof of the Corollary \ref{linearoncomplement}}
%\begin{figure*}
%\textit{Proof:}
For any $\textbf{x}, \textbf{y}  \in R^n$, $\textbf{x}=(x_1\ x_2\dots x_n)$ and $\textbf{y}=(y_1\ y_2\dots y_n)$.
\begin{equation*}
\begin{split}
Consider \ \phi(a\textbf{x}+b\textbf{y})^c = & (\phi(ax_1+by_1)^c\ \phi(ax_2+by_2)^c\dots \phi(ax_n+by_n)^c). \\ Thus \
    \phi^{-1}(\phi(a\textbf{x}+b\textbf{y})^c) = & (\phi^{-1}(\phi(ax_1+by_1)^c)\ \phi^{-1}(\phi(ax_2+by_2)^c)\dots \phi^{-1}(\phi(ax_n+by_n)^c)) \\
     = & (ax_1+by_1+2+2w\ ax_2+by_2+2+2w\dots ax_n+by_n+2+2w) \\
     = & (a(x_1+2+2w)+b(y_1+2+2w)\ a(x_2+2+2w)+b(y_2+2+2w)\dots \\
        & a(x_n+2+2w)+b(y_n+2+2w)) \\
     = & (a\phi^{-1}(\phi(x_1)^c)+b\phi^{-1}(\phi(y_1)^c)\ a\phi^{-1}(\phi(x_2)^c)+b\phi^{-1}(\phi(y_2)^c)\dots \\
        & \ldots a\phi^{-1}(\phi(x_n)^c)+b\phi^{-1}(\phi(y_n)^c)) \\
     = & a((\phi^{-1}(\phi(x_n)^c)\ \phi^{-1}(\phi(x_{n-1})^c)\ \dots \phi^{-1}(\phi(x_1)^c)) \\ 
        & + b(\phi^{-1}(\phi(y_n)^c)\ \phi^{-1}(\phi(y_{n-1})^c)\dots\phi^{-1}(\phi(y_1)^c)) \\
     = & a\phi^{-1}(\phi(\textbf{x})^c) + b\phi^{-1}(\phi(\textbf{y})^c).
\end{split}
\end{equation*}
%\end{figure*}

\end{document}